\documentclass[reqno]{amsart}

\usepackage{amsmath, amssymb}
\usepackage{graphicx}
\usepackage{dcolumn}
\usepackage{bm}
\usepackage{hyperref}
\usepackage{setspace}
\usepackage{color}

\title{%
  Constructing of constraint preserving scheme for Einstein equations
}%

\author[T. Tsuchiya]{Takuya Tsuchiya}
\author[G. Yoneda]{Gen Yoneda}
\address[T. Tsuchiya, G. Yoneda]{
  Department of Mathematics, School of Fundamental Science and Engineering,
  Waseda University, Okubo, Shinjuku, Tokyo 169-8555, Japan}
\email[T. Tsuchiya]{t-tsuchiya@aoni.waseda.jp, tsuchiya@akane.waseda.jp}

\begin{document}

\maketitle


\begin{abstract}
  We propose a new numerical scheme of evolution for the  Einstein equations
  using the discrete variational derivative method
  (DVDM).
  We derive the discrete evolution equation of the constraint using this scheme
  and show the constraint preserves in the discrete level.
  In addition, to confirm the numerical stability using this scheme, we perform
  some numerical simulations by discretized equations with the Crank-Nicolson
  scheme and with the new scheme, and we find that the new discretized equations
  have better stability than that of the Crank-Nicolson scheme.
\end{abstract}

\section{Introduction}

Recently, the gravitational waves which are the solutions of the Einstein
equations and are the part of the most interesting astrophysical events were
observed by LIGO \cite{AbbottGW09,AbbottGW12}.
Since the Einstein equations are sets of the nonlinear partial differential
equations with constraints, numerical calculations are the realistic tools to
study these equations for the actual physical events in the universe such as the
gravitational waves.
Analyzing the Einstein equations numerically is called ``numerical relativity'',
and these studies made a substantial contribution to observe the
gravitational wave events.
These discoveries must open new physical fields to survey of lots of
astrophysical events.
Therefore, it is important to improve numerical techniques to make high accuracy
numerical results.

Although, there are some researches of the suitable numerical
schemes of the Einstein equations(e.g. \cite{RL08}), it is not enough to perform
precisely simulations.
However, it would be hardly to make best discretized equations of the Einstein
equations because of the strong nonlinearity.
In addition, some of the characteristics in the continuous system are often lost
in the process of transforming the continuous equation to the discretized one.
There are some discretized schemes of preserving properties of the continuous
equations.
One of the schemes is the discrete variational derivative method (DVDM).
DVDM was proposed and has been extended by Furihata, Mori, Matsuo, and Yaguchi
\cite{FM96, Furihata99, MF01,
 FM10, AMM15, IY15}, and this method is one of the
structure-preserving methods \cite{HC10}.
The method of formulating a discrete system using DVDM is similar to the
variational principle.
In the variational principle, we provide a function such as the
Lagrangian density and can obtain a continuous system using the functional
derivative of the function.
On the other hand, in DVDM, we provide a discretized function of the Lagrangian
density and can obtain a discrete system using the discrete variational
derivative.
The Crank-Nicolson scheme (CNS) and the Runge-Kutta scheme are general methods
of discretizing equations, and they often use in numerical relativity.
On the other hand, DVDM is a specific method of preserving constraints and
diffusion characteristics in the continuous system.
Since the Einstein equations include the constraints, DVDM can provide a
discretized Einstein equations with constraint preserving properties.

Before we try to make the discretized Einstein equations using DVDM, we studied
the discretized Maxwell's equations using the method because of the complexities
of the Einstein equations\cite{TY16}.
Since the Maxwell's equations are classified the hyperbolic equations with
constraints, the characters of the system are similar to those of the Einstein
equations.
In \cite{TY16}, the authors studied the numerical stabilities for the Maxwell's
equations using DVDM, and showed that DVDM is effective to perform stable
simulations in the system.

In this Letter, indices such as $(i, j, k, \cdots)$ run from $1$ to $3$.
We use the Einstein convention of summation of repeated up-down indices.

\section{DVDM}

Now, we review the processes of the standard method of discretized equations
using general schemes and the new way using DVDM.
The discrete values of the variable $u$ are defined as $u{}^{(n)}_{(k)}$.
The upper index $(n)$ and the lower index $(k)$ denote the time components and
the space components, respectively.
The details of this are in Ref. \cite{FM10}.
We adopt only the canonical formulation in this Letter, so that these processes
are only shown in the case of the Hamiltonian density.
First, we introduce the process of the general case of constructing the discrete
equations such as CNS and the Runge-Kutta scheme.
(A): We set a Hamiltonian density ${H}(q_i, p^i)$ with the variable $q_i$ and
the canonical conjugate momentum $p^i$.
(B): The dynamical equations are given as
\begin{align}
  \dot{q}_i&=\frac{\delta {H}}{\delta p^i},\quad
  \dot{p}^i=-\frac{\delta {H}}{\delta q_i},
\end{align}
then we discretize the continuous equations following the ideas of the schemes
such as CNS and the Runge-Kutta scheme.
Next, the processes of discrete equations using DVDM are as follows:
(A$'$): We set a discrete Hamiltonian density ${H}^{(n)}_{(k)}\bigl(
q_i{}^{(n)}_{(k)}, p^i{}^{(n)}_{(k)}\bigr)$ with the discrete variable
$q_i{}^{(n)}_{(k)}$ and the discrete canonical conjugate momentum
$p^i{}^{(n)}_{(k)}$.
(B$'$): The calculation of the discretized system using DVDM is
\begin{align}
  \frac{q_i{}^{(n+1)}_{(k)}-q_{i}{}^{(n)}_{(k)}}{\Delta t}
  &=\frac{\widehat{\delta}{H}}{\widehat{\delta}(p^i{}^{(n+1)}_{(k)},
    p^i{}^{(n)}_{(k)})},\\
  \frac{p^i{}^{(n+1)}_{(k)}-p^{i}{}^{(n)}_{(k)}}{\Delta t}
  &=-\frac{\widehat{\delta}{H}}{\widehat{\delta}(q_i{}^{(n+1)}_{(k)},
    q_i{}^{(n)}_{(k)})},
\end{align}
where $\widehat{\delta}{H}/\bigl(\widehat{\delta}(p^i{}^{(n+1)}_{(k)},
p^i{}^{(n)}_{(k)})\bigr)$ and $\widehat{\delta} {H}/\bigl(
\widehat{\delta}(q_i{}^{(n+1)}_{(k)},q_i{}^{(n)}_{(k)})\bigr)$ are calculated as
\begin{align}
  {H}^{(n+1)}_{(k)} - {H}^{(n)}_{(k)}
  &=
  \frac{\widehat{\delta} \mathcal{H}}{\widehat{\delta}(p^i{}^{(n+1)}_{(k)},
    p^i{}^{(n)}_{(k)})}(p^{i}{}^{(n+1)}_{(k)}-p^{i}{}^{(n)}_{(k)})
  \nonumber\\
  &+ \frac{\widehat{\delta} {H}}{\widehat{\delta}(q_i{}^{(n+1)}_{(k)},
    q_i{}^{(n)}_{(k)})}(q_{i}{}^{(n+1)}_{(k)}-q_{i}{}^{(n)}_{(k)}).
\end{align}

\section{Densitized ADM formulation of only $t$}

The Lagrangian density of the Einstein equations is given by
\begin{align}
  \mathcal{L}^{\text{GR}}=\alpha\sqrt{\gamma}(R-K^2+K_{ij}K^{ij}),
  \label{eq:Lagrangian}
\end{align}
where $\alpha$ is the lapse function, $\gamma$ is the determinant of the
three-metric $\gamma_{ij}$, $R$ is the three-scalar curvature,
$K_{ij}$ is the extrinsic curvature, and $K=\gamma^{ij}K_{ij}$.
The word GR means the General Relativity.
With the variable $\gamma_{ij}$ and the canonical momentum $\pi^{ij} \equiv
(\delta\mathcal{L}^{\text{GR}})/(\delta\partial_t\gamma_{ij})$, the Hamiltonian
formulation of the Einstein equations can be constructed.
This formulation was given by Arnowitt, Deser, and Misner (ADM) \cite{ADM62, Wald84}.
This is the first successful decomposition of the Einstein equations to space
and time.

If we discretize the equations of the formulation with DVDM, we face to two
difficulties.
First is discretizing  of $\gamma$.
The variational of $\gamma$ in continuous case is
\begin{align}
  \delta\gamma = \gamma \gamma^{ij}\delta\gamma_{ij}.
\end{align}
It is hard to make the above relation in discretized case.
Then,
we use the conformal transformation such as
\begin{align}
  \gamma_{ij}=\gamma\tilde{\gamma}_{ij},
  K_{ij}=\gamma^{1/2}\tilde{K}_{ij},
  \alpha=\gamma^{1/2}\tilde{\alpha},
  \label{eq:ConformalTrans}
\end{align}
to vanish $\gamma$ in the Lagrangian density.
These relations are derived from following two conditions:
(I) the Lagrangian density $\mathcal{L}^{\text{GR}}$ does not include
$\sqrt{\gamma}$, (II) the evolution equation of $\tilde{\gamma}_{ij}$ does not
include $\sqrt{\gamma}$.
If these conditions are satisfied, the $\sqrt{\gamma}$ is not also included in
the expression of the Hamiltonian density $\tilde{\mathcal{H}}^{\text{GR}}$
derived from the Legendre transformation of the Lagrangian density.
Thus, the canonical formulation using $\tilde{\mathcal{H}}^{\text{GR}}$ does not
include $\sqrt{\gamma}$.
With \eqref{eq:ConformalTrans}, the Lagrangian density \eqref{eq:Lagrangian} can
be expressed as
\begin{align}
  \tilde{\mathcal{L}}^{\text{GR}}
  &= \tilde{\alpha}
  \biggl\{\tilde{R}
  + \tilde{\gamma}^{ij}(\partial_i\tilde{\Gamma}^m{}_{mj})
  - \tilde{\Gamma}^m{}_{mi}\tilde{\gamma}^{ab}\tilde{\Gamma}^i{}_{ab}
  - \frac{1}{8}\tilde{\Gamma}^m{}_{mi}\tilde{\Gamma}^n{}_{nj}
  \tilde{\gamma}^{ij}
  - \tilde{\pi}^2 + \tilde{\pi}^{ij}\tilde{\pi}_{ij}
  \biggr\},
\end{align}
where $\tilde{R}$ is the scalar curvature of $\tilde{\gamma}_{ij}$,
$\tilde{\Gamma}^\ell{}_{ij}$ is the connection of $\tilde{\gamma}_{ij}$,
$\tilde{\pi}^{ij}$ is the canonical momentum of $\tilde{\gamma}_{ij}$,
and $\tilde{\pi}\equiv
\tilde{\gamma}_{ij}\tilde{\pi}^{ij}$.
Note that the expressions of $\mathcal{L}^{\text{GR}}$ and
$\tilde{\mathcal{L}}^{\text{GR}}$ are different, but they are same, that is to
say $\mathcal{L}^{\text{GR}}=\tilde{\mathcal{L}}^{\text{GR}}$.
The relation can be derived in straightforward.
Then the Hamiltonian density is defined as
\begin{align}
  \tilde{\mathcal{H}}^{\text{GR}}
  &\equiv \tilde{\pi}^{ij}(\partial_t\tilde{\gamma}_{ij})
  - \tilde{\mathcal{L}}^{\text{GR}}.
\end{align}
Next is construction of suitable discretization of the $R$.
This is defined as in continuous case such as
\begin{align}
  R
  &= \frac{1}{2}\gamma^{ij}(\partial_\ell\Gamma^\ell{}_{ij}
  - \partial_i\Gamma^\ell{}_{i\ell}
  + \Gamma^{\ell}{}_{m\ell}\Gamma^m{}_{ij}
  - \Gamma^{\ell}{}_{mi}\Gamma^m{}_{j\ell}),\\
  \Gamma^{\ell}{}_{ij}
  &\equiv
  \frac{1}{2}\gamma^{\ell m}(\partial_i\gamma_{m j}
  + \partial_j\gamma_{m i}
  - \partial_m\gamma_{ij}).
\end{align}
Note that there are the same relations between $\tilde{R}$,
$\tilde{\Gamma}^\ell{}_{ij}$, and $\tilde{\gamma}_{ij}$.
This means that the conformal transformation \eqref{eq:ConformalTrans} does not
overcome the difficulty of $R$.
Therefore we treat only $t$ to simplify $R$ in this Letter.

If $\tilde{\mathcal{H}}^{\text{GR}}$ is the function of only $t$, the density
can be expressed as
\begin{align}
  \tilde{\mathcal{H}}^{\text{GR}}
  &= \tilde{\alpha}(-\tilde{\pi}^2 + \tilde{\pi}^{ij}\tilde{\pi}_{ij}).
\end{align}
Then the evolution equations are
\begin{align}
  \partial_t\tilde{\gamma}_{ij}
  &\equiv \frac{\delta\tilde{\mathcal{H}}^{\text{GR}}}{
    \delta\tilde{\pi}^{ij}}
  = -2\tilde{\alpha}\tilde{\pi}\tilde{\gamma}_{ij}
  + 2\tilde{\alpha}\tilde{\pi}_{ij},\\
  \partial_t\tilde{\pi}^{ij}
  &\equiv -\frac{\delta\tilde{\mathcal{H}}^{\text{GR}}}{
    \delta\tilde{\gamma}_{ij}}
  = 2\tilde{\alpha}\tilde{\pi}\tilde{\pi}^{ij}
  - 2\tilde{\alpha}\tilde{\pi}^{mi}\tilde{\pi}_m{}^j,
\end{align}
and the constraint equation is
\begin{align}
  \tilde{\mathcal{H}}
  &\equiv -\frac{\delta\tilde{\mathcal{H}}^{\text{GR}}}{
    \delta\tilde{\alpha}}
  = \tilde{\pi}^2 - \tilde{\pi}^{ij}\tilde{\pi}_{ij}.
  \label{eq:HamiltonianConstraint}
\end{align}
In the ADM formulation, the constraint values are the Hamiltonian constraint
$\mathcal{H}$ and the momentum constraints  $\mathcal{M}_i$, thus there are
four constraints.
However, there is only one equation of constraint in the above set of equations.
This is because we treat the case of $\tilde{\mathcal{H}}^{\text{GR}}
= \tilde{\mathcal{H}}^{\text{GR}}(t)$.
If we treat the case of $\tilde{\mathcal{H}}=\tilde{\mathcal{H}}^{\text{GR}}(t,x,y,z)$, there are
four constraint equations in the new formulation.

The evolution equation of \eqref{eq:HamiltonianConstraint} can be derived as
\begin{align}
  \partial_t\tilde{\mathcal{H}}
  &=
  2\tilde{\pi}\tilde{\pi}^{ij}(\partial_t\tilde{\gamma}_{ij})
  + 2\tilde{\pi}\tilde{\gamma}_{ij}(\partial_t\tilde{\pi}^{ij})
  - 2\tilde{\pi}^{ij}\tilde{\pi}_{i}{}^m
  (\partial_t\tilde{\gamma}_{jm})
  - 2\tilde{\pi}_{ij}(\partial_t\tilde{\pi}^{ij})
  \nonumber\\
  &=
  2\tilde{\pi}\tilde{\pi}^{ij}(-2\tilde{\alpha}\tilde{\pi}\tilde{\gamma}_{ij}
  + 2\tilde{\alpha}\tilde{\pi}_{ij})
  + 2\tilde{\pi}\tilde{\gamma}_{ij}
  \left(2\tilde{\alpha}\tilde{\pi}\tilde{\pi}^{ij}
  - 2\tilde{\alpha}\tilde{\pi}_{m}{}^i\tilde{\pi}^{mj}
  \right)
  \nonumber\\
  &\quad
  - 2\tilde{\pi}^{ij}\tilde{\pi}_{i}{}^m
  (-2\tilde{\alpha}\tilde{\pi}\tilde{\gamma}_{jm}
  + 2\tilde{\alpha}\tilde{\pi}_{jm})
  - 2\tilde{\pi}_{ij}\left(2\tilde{\alpha}\tilde{\pi}\tilde{\pi}^{ij}
  - 2\tilde{\alpha}\tilde{\pi}_{m}{}^i\tilde{\pi}^{mj}
  \right)\nonumber\\
  &=0,
\end{align}
then the Hamiltonian constraint is satisfied in the evolutions.
We call the evolution equation of constraint as constraint propagation (CP).

\section{Discretized equations
  \label{sec:discretizedEqs}
}

\subsection{DVDM}

We set the discrete Hamiltonian density as
\begin{align}
  \tilde{\mathcal{H}}^{\text{GR}}{}^{(n)}
  &\equiv
  \tilde{\alpha}^{(n)}\left(-(\tilde{\pi}^{(n)})^2
  + \tilde{\pi}^{i}{}_j{}^{(n)}\tilde{\pi}^{j}{}_i{}^{(n)}\right),
\end{align}
then, the discretized evolution equations with DVDM are
\begin{align}
  \frac{\tilde{\gamma}_{ij}^{(n+1)}
    - \tilde{\gamma}_{ij}{}^{(n)}}{\Delta t}
  &=
  \tilde{\alpha}^{(n+1)}
  \biggl\{-(\tilde{\pi}^{(n+1)} + \tilde{\pi}^{(n)})
  \tilde{\gamma}_{ij}{}^{(n)}
  + \frac{1}{2}
  (\tilde{\pi}^\ell{}_i{}^{(n+1)}
  + \tilde{\pi}^\ell{}_i{}^{(n)})
  \tilde{\gamma}_{j\ell}{}^{(n)}
  \nonumber\\
  &\quad
  + \frac{1}{2}
  (\tilde{\pi}^\ell{}_j{}^{(n+1)}
  + \tilde{\pi}^\ell{}_j{}^{(n)})
  \tilde{\gamma}_{i\ell}{}^{(n)}
  \biggr\},
  \label{eq:GammaEvoDVDM}\\
  \frac{\tilde{\pi}^{ij}{}^{(n+1)}
    - \tilde{\pi}^{ij}{}^{(n)}}{\Delta t}
  &=
  \tilde{\alpha}^{(n+1)}\biggl\{
  (\tilde{\pi}^{(n+1)} + \tilde{\pi}^{(n)})
  \tilde{\pi}^{ij}{}^{(n+1)}
  - \frac{1}{2}
  (\tilde{\pi}^{i}{}_\ell{}^{(n+1)}+\tilde{\pi}^i{}_\ell{}^{(n)})
  \tilde{\pi}^{\ell j}{}^{(n+1)}
  \nonumber\\
  &\quad
  - \frac{1}{2}
  (\tilde{\pi}^{j}{}_\ell{}^{(n+1)}+\tilde{\pi}^j{}_\ell{}^{(n)})
  \tilde{\pi}^{\ell i}{}^{(n+1)}
  \biggr\}.
  \label{eq:PiEvoDVDM}
\end{align}
The above equations, \eqref{eq:GammaEvoDVDM}-\eqref{eq:PiEvoDVDM},
include asymmetric time components in each r.h.s. to satisfy the discretized
Hamiltonian constraint in the evolutions.
Actually, the discretized CP is derived as
\begin{align}
  &\quad\frac{\tilde{\mathcal{H}}^{(n+1)} - \tilde{\mathcal{H}}^{(n)}}{\Delta t}
  \nonumber\\
  &=
  (\tilde{\pi}^{(n+1)} + \tilde{\pi}^{(n)})\tilde{\pi}^{ij}{}^{(n+1)}
  \frac{\tilde{\gamma}_{ij}{}^{(n+1)} - \tilde{\gamma}_{ij}{}^{(n)}}{\Delta t}
  \nonumber\\
  &\quad
  + (\tilde{\pi}^{(n+1)} + \tilde{\pi}^{(n)})\tilde{\gamma}_{ij}{}^{(n)}
  \frac{\tilde{\pi}^{ij}{}^{(n+1)} - \tilde{\pi}^{ij}{}^{(n)}}{\Delta t}
  \nonumber\\
  &\quad
  - (\tilde{\pi}^i{}_\ell{}^{(n+1)} + \tilde{\pi}^i{}_\ell{}^{(n)})
  \tilde{\pi}^{\ell j}{}^{(n+1)}
  \frac{\tilde{\gamma}_{ij}{}^{(n+1)} - \tilde{\gamma}_{ij}{}^{(n)}}{\Delta t}
  \nonumber\\
  &\quad
  - (\tilde{\pi}^\ell{}_j{}^{(n+1)} + \tilde{\pi}^\ell{}_j{}^{(n)})
  \tilde{\gamma}_{i\ell}{}^{(n)}
  \frac{\tilde{\pi}^{ij}{}^{(n+1)} - \tilde{\pi}^{ij}{}^{(n)}}{\Delta t}
  \nonumber\\
  &= 0.
\end{align}
Thus, the discretized CP of DVDM is the same as the continuous case.

\subsection{Crank-Nicolson scheme}

To compare with the numerical stability of the discrete system using DVDM,
we select the Crank-Nicolson scheme (CNS) which is used in the numerical
relativity.
The discrete evolution equations with CNS are
\begin{align}
  &\quad
  \frac{\tilde{\gamma}_{ij}{}^{(n+1)} - \tilde{\gamma}_{ij}{}^{(n)}}{\Delta t}
  \nonumber\\
  &= -\frac{1}{8}(\tilde{\alpha}^{(n+1)} + \tilde{\alpha}^{(n)})
  (\tilde{\pi}^{mn}{}^{(n+1)} + \tilde{\pi}^{mn}{}^{(n)})
  (\tilde{\gamma}_{mn}{}^{(n+1)} + \tilde{\gamma}_{mn}{}^{(n)})
  (\tilde{\gamma}_{ij}{}^{(n+1)} + \tilde{\gamma}_{ij}{}^{(n)})
  \nonumber
  \\
  &\quad+ \frac{1}{8}(\tilde{\alpha}^{(n+1)} + \tilde{\alpha}^{(n)})
  (\tilde{\pi}^{mn}{}^{(n+1)} + \tilde{\pi}^{mn}{}^{(n)})
  (\tilde{\gamma}_{mi}{}^{(n+1)} + \tilde{\gamma}_{mi}{}^{(n)})
  (\tilde{\gamma}_{nj}{}^{(n+1)} + \tilde{\gamma}_{nj}{}^{(n)}),
  \label{eq:GammaEvoCN}\\
  &\quad
  \frac{\tilde{\pi}^{ij}{}^{(n+1)} - \tilde{\pi}^{ij}{}^{(n)}}{\Delta t}
  \nonumber\\
  &= \frac{1}{8}(\tilde{\alpha}^{(n+1)}+ \tilde{\alpha}^{(n)})
  (\tilde{\pi}^{mn}{}^{(n+1)} + \tilde{\pi}^{mn}{}^{(n)})
  (\tilde{\gamma}_{mn}{}^{(n+1)} + \tilde{\gamma}_{mn}{}^{(n)})
  (\tilde{\pi}^{ij}{}^{(n+1)} + \tilde{\pi}^{ij}{}^{(n)})
  \nonumber\\
  &\quad
  - \frac{1}{8}(\tilde{\alpha}^{(n+1)}+\tilde{\alpha}^{(n)})
  (\tilde{\pi}^{mi}{}^{(n+1)}+\tilde{\pi}^{mi}{}^{(n)})
  (\tilde{\pi}^{nj}{}^{(n+1)} + \tilde{\pi}^{nj}{}^{(n)})
  (\tilde{\gamma}_{mn}{}^{(n+1)} + \tilde{\gamma}_{mn}{}^{(n)}),
  \label{eq:PiEvoCN}
\end{align}
and the discretized CP with the scheme is
\begin{align}
  &\quad\frac{\tilde{\mathcal{H}}^{(n+1)} - \tilde{\mathcal{H}}^{(n)}}{\Delta t}
  \nonumber\\
  &=
  \frac{1}{8}(\tilde{\alpha}^{(n+1)} + \tilde{\alpha}^{(n)})
  (\tilde{\pi}^{mn}{}^{(n+1)} + \tilde{\pi}^{mn}{}^{(n)})
  \biggl\{
  \nonumber\\
  &\quad
  - (\tilde{\pi}^{(n+1)} + \tilde{\pi}^{(n)})
  (\tilde{\pi}{}^{(n+1)}-\tilde{\pi}^{(n)})
  (\tilde{\gamma}_{mn}{}^{(n+1)} + \tilde{\gamma}_{mn}{}^{(n)})
  \nonumber\\
  &\quad
  + (\tilde{\pi}^{(n+1)} + \tilde{\pi}^{(n)})
  (\tilde{\pi}^{i}{}_n{}^{(n+1)} - \tilde{\pi}^{i}{}_n{}^{(n)})
  (\tilde{\gamma}_{mi}{}^{(n+1)} + \tilde{\gamma}_{mi}{}^{(n)})
  \nonumber\\
  &\quad
  +(\tilde{\pi}^i{}_\ell{}^{(n+1)} + \tilde{\pi}^i{}_\ell{}^{(n)})
  (\tilde{\pi}^{\ell}{}_i{}^{(n+1)}-\tilde{\pi}^{\ell}{}_i{}^{(n)})
  (\tilde{\gamma}_{mn}{}^{(n+1)} + \tilde{\gamma}_{mn}{}^{(n)})
  \nonumber\\
  &\quad
  - (\tilde{\pi}^i{}_\ell{}^{(n+1)} + \tilde{\pi}^i{}_\ell{}^{(n)})
  (\tilde{\gamma}_{nj}{}^{(n+1)} + \tilde{\gamma}_{nj}{}^{(n)})
  (\tilde{\pi}^{\ell j}{}^{(n+1)}\tilde{\gamma}_{mi}{}^{(n+1)}
  - \tilde{\pi}^{\ell j}{}^{(n)}\tilde{\gamma}_{mi}{}^{(n)})
  \biggr\}.
\end{align}
This is not equal to zero in general because that the terms
$(\tilde{\pi}{}^{(n+1)}-\tilde{\pi}^{(n)})$,
$(\tilde{\pi}^{i}{}_n{}^{(n+1)} - \tilde{\pi}^{i}{}_n{}^{(n)})$,
$(\tilde{\pi}^{\ell}{}_i{}^{(n+1)}-\tilde{\pi}^{\ell}{}_i{}^{(n)})$,
and $(\tilde{\pi}^{\ell j}{}^{(n+1)}\tilde{\gamma}_{mi}{}^{(n+1)}
- \tilde{\pi}^{\ell j}{}^{(n)}\tilde{\gamma}_{mi}{}^{(n)})$
are not equal to zero.
Therefore, the discrete constraint value of CNS would not be conserved
in the numerical simulations.

\section{Numerical tests}

We perform some simulations to study differences of the constraint violations
using the two schemes.
Then, we adopt the initial condition as
\begin{align}
  ds^2=-dt^2+t^{-4/7}dx^2+t^{6/7}dy^2+t^{12/7}dz^2.
\end{align}
This is the Kasner solution which is one of the exact solutions of the Einstein
equations.
This solution expresses the expanding universe, and there is a singularity at
$t=0$.
Then the initial conditions of $\alpha$, $\gamma_{ij}$, and $K_{ij}$ are given by
\begin{align}
  \left\{
  \begin{array}{l}
    \alpha=1,\\
    \gamma_{ij}=\text{diag}(t^{-4/7},t^{6/7},t^{12/7}),\\
    K_{ij}
    =\text{diag}((4/7)t^{-11/7},-(6/7)t^{-1/7},-(12/7)t^{5/7}),
  \end{array}
  \right.
  \label{eq:initialData1}
\end{align}
and \eqref{eq:initialData1} using the relations of \eqref{eq:ConformalTrans}
becomes as
\begin{align}
  &\left\{
  \begin{array}{l}
    \tilde{\alpha}=t^{-1},
    \\
    \tilde{\gamma}_{ij}=\text{diag}(t^{-18/7},t^{-8/7},t^{-2/7}),\\
    \tilde{\pi}^{ij}
    =\text{diag}(-(2/7)t^{-54/7},(3/7)t^{-24/7},(6/7)t^{-6/7}).
    \label{eq:Kasner}
  \end{array}
  \right.
\end{align}
In addition, it is hardly to make constraint violations in the above condition.
Then, we add perturbations such as
\begin{align}
  \gamma_{11} = t^{-4/7} + 0.05.
\end{align}
We set the other conditions in these simulations such as
\begin{itemize}
\item initial time: $t=20$.
\item time steps: $dt=0.25$.
\item $\alpha$(gauge condition) : exact solution of \eqref{eq:Kasner}.
\item boundary condition: periodic.
\end{itemize}
There are $(n+1)$ time step variables in the discretized evolution equations
using DVDM and CNS (\eqref{eq:GammaEvoDVDM}-\eqref{eq:PiEvoDVDM} and
\eqref{eq:GammaEvoCN}-\eqref{eq:PiEvoCN}, respectively).
In the simulations, the variables are calculated in iterative because of the
nonlinearity, then we set the numbers of times of iteration as
fourth order in both schemes.
\begin{figure}[htbp]
  \centering
  \includegraphics[keepaspectratio=true,width=0.7\hsize]{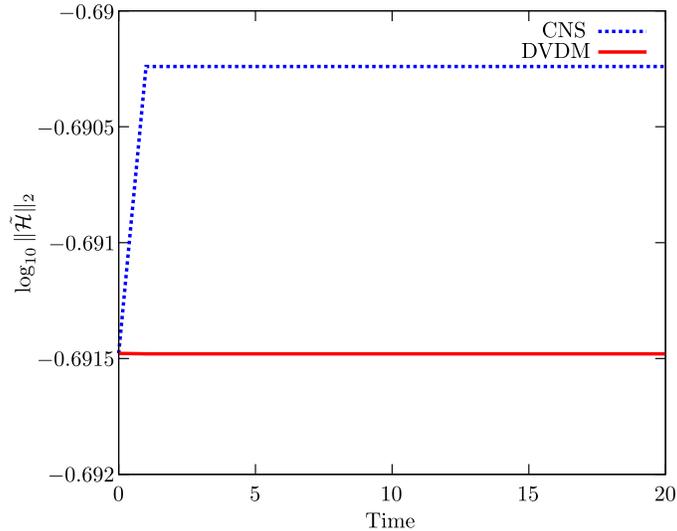}
  \caption{Vertical axis means the violations of $\tilde{\mathcal{H}}$ and
    horizontal axis means time.
    The solid line is drawn by DVDM and the dotted line is by CNS.
    The line of CNS is grow up after initial time.
    On the other hand, the one of DVDM is horizontal in the simulations.
    \label{fig:Hamiltonain}}
\end{figure}
Fig.\ref{fig:Hamiltonain} is the constraint violations using DVDM and CNS.
In this Figure, the solid line which is drawn by DVDM is horizontal, and the
dotted line which is drawn by CNS is rising.
This means that the values of the constraint $\tilde{\mathcal{H}}$ of DVDM is
unchanged in the evolution.
On the other hand, that of CNS is increasing from the initial values.
Therefore, these results are consistent with the results of analyzing CP in
Sec.\ref{sec:discretizedEqs}.

\section{Summary}

We derived suitable discretized equations of the Einstein equations using the
discrete variational derivative method (DVDM).
In the process of the discretization, there are two difficulties.
One is the existence of the determinant of $\gamma_{ij}$, $\gamma$, in the
Lagrangian density, and the other is the complexity of the scalar curvature
${}^{(3)}R$ included in the density.
For the existence of $\gamma$, we eliminated it by the conformal transformation
of the variables.
For the complexity of ${}^{(3)}R$, we restricted the coordinate as only $t$.
Then, we derived the discretized evolution equations of the variables and the
constraint in DVDM and the Crank-Nicolson scheme (CNS), and showed the discrete
constraint value of DVDM is conserved and the one of CNS is not conserved.
In addition, we performed some simulations by using DVDM and CNS.
Then, the constraint value using DVDM was conserved.
On the other hand, the value using CNS was not conserved.
These results are consistent with the analysis of the discretized evolution
equations of the constraint using DVDM and CNS.
Therefore DVDM is the validity to perform stable simulations in the
numerical relativity.

In this Letter, we restricted the coordinates to simplify the complexities of
the scalar curvature $R$.
To perform realistic simulations, we should handle the $R$ in the coordinates
$(t,x,y,z)$.
In future work, we would like to study in the coordinates.

\section*{Acknowledgements}
G. Y. was partially supported by a Waseda University Grant for Special
Research Projects (number 2016B-121, 2016K-161).

\end{document}